\documentclass{iucrjournals}
\usepackage{amsmath}

\title{Accurate simulations of magnetic excitations in the neutron simulation package McStas}
\usepackage{algorithm}
\usepackage{algpseudocode}
\usepackage{makecell}
\usepackage{subcaption}
\usepackage{svg}
\usepackage{hyperref}
\usepackage{pdfpages}

\author[a]{Silas B. Schack\IUCrOrcidlink{0009-0001-8057-7859}}%
\author[a]{Kristine M. L. Krighaar\IUCrOrcidlink{0000-0003-1415-5815
}}
\author[a,b]{Emma Y. Lenander\IUCrOrcidlink{0000-0002-1274-3579}}
\author[a]{Kim Lefmann\IUCrCemaillink{lefmann@nbi.ku.dk}\IUCrOrcidlink{0000-0003-4282-756X}}%

\affil[a]{Nanoscience Center, Niels Bohr Institute, University of Copenhagen, Denmark}
\affil[b]{Institute  of  Physics,  \'Ecole  Polytechnique  F\'ed\'erale  de  Lausanne  (EPFL),  CH-1015  Lausanne,  Switzerland}

\begin{document} 
\maketitle 

\begin{synopsis}
We present an accurate algorithm for simulating neutron scattering from spin waves in the ray-tracing package McStas.

\end{synopsis}

\begin{abstract}

A new component for the accurate simulation of neutron scattering from magnetic excitations has been developed for the neutron ray-tracing software McStas. The component \verb|SpinWave_BCO| simulates inelastic neutron scattering from ferro-, antiferro-, and altermagnetic excitations in a body-centered orthorhombic crystal structure, where the dispersion relation and scattered neutron intensities are derived using linear spin wave theory.

Data from a simulated Triple-Axis Spectrometer with an extremely high resolution have been verified by direct comparison with theory and by comparison to data simulated using the package SpinW.
The component serves as a proof-of-concept for the implementation of a more general linear spin wave component in McStas.

\end{abstract}

\keywords{Spin wave; Neutron scattering; ray-tracing simulation; McStas; Altermagnetism}

\section{Introduction}
Neutron scattering is an essential tool for investigating the structure and dynamics of condensed matter, being sensitive to both the nuclei and the magnetic moments in a sample \cite{boothroyd_principles_2020}. In a neutron scattering experiment, one generally measures both the neutron energy and momentum transfers:
\begin{eqnarray}
        \hbar \omega &=& E_{\rm i} - E_{\rm f} = \frac{\hbar^2}{2 m_{\rm n}} (k_{\rm i}^2 - k_{\rm f}^2) \\ 
        {\bf q} &=& {\bf k}_{\rm i} - {\bf k}_{\rm f} ,
\end{eqnarray}
where $m_{\rm n}$ is the neutron mass, $E$ is the neutron energy, {\bf k} is the neutron wave vector, and the indices ''i`` and ''f`` denote the initial and final neutron states, respectively.

Since access to neutron scattering beam time is limited, much instrumental design and preparation is carried out by neutron ray-tracing simulations. Here, virtual experiments \cite{lefmann_virtual_2008} on digital twins of neutron instruments can provide valuable insight into instrument performance, experiment feasibility, experiment resolution, multiple scattering effects, and signal-to-background relations. In addition, much simulation work is  performed for the analysis and design of novel neutron instrumentation, {\em e.g.}\ at the upcoming European Spallation Source, ESS \cite{Andersen2020}.

One of the most used ray-tracing tools is McStas, a package originating from the 1990'ies \cite{Lefmann1999} and continuously upgraded since. In the package, neutron instruments are assembled by positioning tailor-made or system-provided components \cite{Willendrup2020}. Among many existing system components, McStas contains a number of model neutron scattering samples, providing incoherent scattering, small-angle scattering, reflectivity, and crystal diffraction \cite{Willendrup2006, Willendrup2021,Christensen2026}. However, the number of inelastic scattering samples in McStas is at present very limited; the only examples being a flat inelastic mode (akin to a crystal electric field level), a single quasielastic Lorentzian line, and a one-branch phonon sample. This seriously limits the breath of inelastic virtual experiments that can be performed in McStas.

A component for magnon scattering, called \verb|Magnon_bcc|, currently exists in the McStas component library and was written by some of us. However, this component is limited in scope and has never been fully developed or validated. 

In this work, we remedy this situation by presenting the new McStas component \verb+SpinWave_BCO+ that calculates and simulates the spin waves of a ferromagnet or a two-sublattice antiferromagnet on the body-centered orthorhombic structure. The component has been tested in a simple virtual experiment using a thermal triple-axis spectrometer with an unrealistically good resolution in both $\hbar\omega$ and {\bf q}. The simulated dispersion and intensity match convincingly both with analytical calculations and the output of the package SpinW \cite{Toth:2015}. The component can be readily upgraded to accept a more general lattice geometry.

\section{Spin wave theory}
We here describe the theory of quantized spin waves (magnons), also known as linear spin wave theory (LSWT), that we have used as a basis for our simulations.
Although this topic has been known for more than half a century, spin wave theory comes in different variations, {\em e.g.}\ how to specify anisotropy constants and infamous factor 2 variations in definitions of the interaction constants. For these reasons, we deem it necessary to document exactly which equations lie behind our simulations. The main focus will be on the antiferromagnetic spin waves on the body-centered orthorhombic structure. However, the component will also simulate ferromagnetic spin waves in the same crystal structure, so a short introduction to the ferromagnet case is also given.

\begin{figure}
    \centering
    \begin{subfigure}{.49\textwidth}
        \centering
        \includegraphics[width=.87\textwidth]{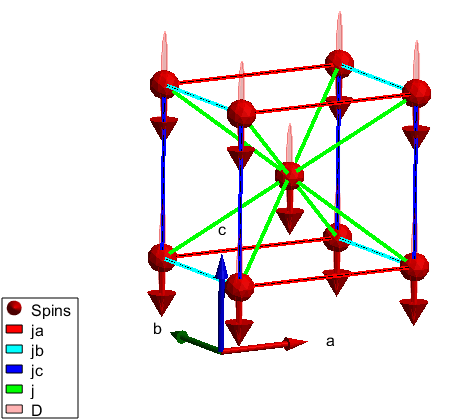} 
    \end{subfigure}
    \begin{subfigure}{.49\textwidth}
        \centering
        \includegraphics[width=.87\textwidth]{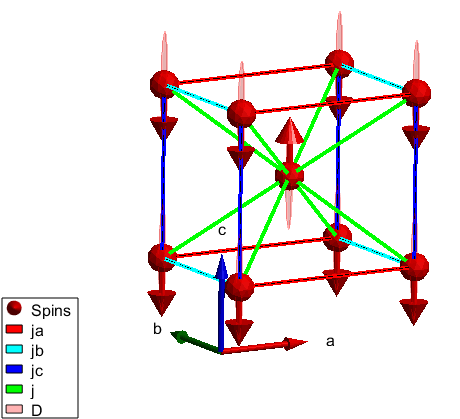}
    \end{subfigure}
     \caption{Magnetic structure and exchange parameters for the ferromagnetic (left) and antiferromagnetic (right) system. The plots are generated using SpinW
     \protect\cite{Toth:2015}.} 
    \label{fig:oI_structure}
\end{figure}
The magnetism in the system of interest is governed by Heisenberg interactions with coupling strength $J_{ij}$, uniaxial anisotropy parametrised by the constant $D$, and an applied magnetic field with field strength $B$. For simplicity, we here align both the easy axis and the field along the crystallographic $c$-axis (the $z$-direction). The field is applied to stabilise the spins in the positive $z$-direction. The system is thus described by the Hamiltonian
\begin{equation}\label{eq:Hamiltonian}
    \mathcal{H}=\frac{1}{2}\sum_{i,j}J_{ij}\mathbf{s}_i\cdot\mathbf{s}_j+\sum_i D(s^z_i)^2-g\mu_BB\sum_is^z_i,
\end{equation}
where $g=2.0023$ is the electron magnetogyric ratio, and $\mu_B=5.7784 \cdot 10^{-2}$~meV/T is the Bohr magneton. When $J_{ij}<0$ dominates, the system is said to be ferromagnetic, while $J_{ij}>0$ dominating leads to an antiferromagnetic system.

\subsection{Ferromagnetic spin waves}
The ferromagnetic case for $D<0$ is relatively simple, as the eigenstate of the Hamiltonian equals the classical ground state, where all spins are oriented along the $+z$ direction.

The magnetic dynamics of the system are found using linear spin wave theory, where the deviations from the ground state are assumed to be small. The Hamiltonian is written using the spin raising and lowering operators for spins at position $\mathbf{m}$: $s_\mathbf{m}^\pm=s^x_\mathbf{m}\pm is_\mathbf{m}^y$. This is then reduced to second order
by the Holstein-Primakoff transformation \cite{boothroyd_principles_2020}:
\begin{alignat}{2}
        \label{eq:ferro_HP}    &s_\mathbf{m}^+=\sqrt{2S}\sqrt{1-a_\mathbf{m}^\dagger a_\mathbf{m}}\cdot a_\mathbf{m}\\
        &s_\mathbf{m}^-=\sqrt{2S}\cdot a_\mathbf{m}^\dagger\sqrt{1-a_\mathbf{m}^\dagger a_\mathbf{m}},\quad\quad \\
        &s_\mathbf{m}^z=S-a^\dagger_\mathbf{m}a_\mathbf{m},
\end{alignat}
where $S$ is (unitless) the spin value, and $a_\mathbf{m}$ and its Hermitian conjugate are bosonic operators satisfying the commutation relation $[a_\mathbf{m},a_\mathbf{m'}^\dagger]=\delta_{\mathbf{m},\mathbf{m'}}$.

Using a first order expansion of the Holstein-Primakoff transformations in the Hamiltonian, and further Fourier transforming the bosonic operators leads directly to the diagonal Hamiltonian
\begin{equation}\label{eq:Hamiltonian_FM}
    \mathcal{H}^{(\rm FM)}=E_0+\sum_{\mathbf{q'}}\hbar\omega_{q'}^{(\rm FM)} a_\mathbf{q'}^\dagger a_\mathbf{q'},
\end{equation}
where $E_0$ is the energy of the ground state and $a_\mathbf{q'}^\dagger a_\mathbf{q'}$ counts the number of magnons of wave vector {\bf q'} and energy $\hbar\omega_{\bf q'}$.
The spin wave dispersion relation is given by 
\begin{eqnarray}\label{eq:FMdispersion}
    \hbar\omega_{q'}^{(\rm FM)}=S(J^{(\rm FM)}(\mathbf{0})-J^{(\rm FM)}(\mathbf{q'}))-\left(S-\frac{1}{2}\right)D+g\mu_BB.
\end{eqnarray}
Here we have defined the Fourier transform of the coupling constants as
\begin{eqnarray}
    J^{(\rm FM)}(\mathbf{q'})=\sum_{\pmb{\delta}} J(\pmb{\delta})e^{i\mathbf{q'}\cdot\pmb{\delta}},
\end{eqnarray}
where the $\pmb{\delta}$'s are vectors connecting one particular spin to its neighbours. We assume couplings $j_a$, $j_b$, and $j_c$ to the nearest neighbours along each crystallographic axis in our body-centered orthorhombic structure. In addition, we add couplings $j$ to the spins in all body centered positions. The system is shown on the left in figure \ref{fig:oI_structure}. It is convenient to separate $J^{(\rm FM)}(\mathbf{q)}$ into a part accounting for the interactions with the body centered spins and spins along the crystallographic axes, respectively. We then obtain
\begin{eqnarray}
    J^{(\rm FM)}(\mathbf{q'})=J(\mathbf{q'})+J_1(\mathbf{q'}),
\end{eqnarray}
with
\begin{align}\label{eq:J_q}
    J(\mathbf{q')}=&8j\cos\left(\frac{aq'_x}{2}\right)\cos\left(\frac{bq'_y}{2}\right)\cos\left(\frac{cq'_z}{2}\right)
\end{align}
and
\begin{equation}\label{eq:J1_q}
    J_1(\mathbf{q'})=2(j_a\cos(aq'_x)+j_b\cos(bq'_y)+j_c\cos(cq'_z)).
\end{equation}

Neutron scattering from magnons happens at low temperatures primarily via the creation or annihilation of a single magnon. The partial differential cross section for unpolarised neutron scattering from single magnons in the ferromagnet is \cite{Marshall:1971}:
\begin{align}\label{eq:inel_cross_FM}
    \left(\frac{d^2\sigma}{d\Omega dE_f}\right)_{\pm}=&(\gamma r_0)^2\Big[\frac{g}{2}F(\mathbf{q})\Big]^2e^{-2W}\frac{k_{\rm f}}{k_{\rm i}}(1+\hat{q}_z^2)\frac{(2\pi)^3}{2v_0}S\nonumber\\&\times \sum_{\mathbf{q'},\pmb{\tau}} \left(n_{\mathbf{q'},a}+\frac{1}{2}\pm \frac{1}{2}\right)\delta(\mathbf{q}\mp\mathbf{q'}-\pmb{\tau})\delta(\hbar\omega\mp\hbar   \omega^{(\rm FM)}_{\mathbf{q'}}),
\end{align}
where $+$ ($-$) refers to the creation (annihilation) of a magnon with energy $\hbar\omega^{(\rm FM)}_\mathbf{q'}$. In the expression, $\gamma = 1.9130$ is the neutron magnetogyric ratio, $r_0 = 2.8179$~fm is the classical electron radius, $\hat{q}_z$ is the $z$-component of a unit vector along the neutron scattering vector {\bf q}, $v_0$ is the volume of the magnetic unit cell, and $n_{\mathbf{q'}}=1/(e^{\hbar\omega^{(\rm FM)}_{\mathbf{q'}}/(k_bT)}-1)$ is the thermal occupation number of (bosonic) magnons. Both the magnetic form factor, $F(\mathbf{q})$, and the Debye-Waller factor, $e^{-2W}$, will be approximated as constants equal to $1$.

\subsection{Antiferromagnetic spin waves}
 In this section we describe the dynamics of spin waves in a classical two-sublattice antiferromagnet (the Néel state). The magnetic unit cell of such an antiferromagnetic system is shown on the right in figure \ref{fig:oI_structure}.

As with the ferromagnet, the magnetic dynamics are found using linear spin wave theory. In this case we perform the Holstein-Primakoff transformations for each sublattice, where the $t_\mathbf{n}$-operators represented the spin operators of the "down" sublattice rotated by 180 degrees about the $x$-axis:
\begin{alignat}{2}
        \label{eq:antiferro_HP}    &s_\mathbf{m}^+=\sqrt{2S}\sqrt{1-a_\mathbf{m}^\dagger a_\mathbf{m}}\cdot a_\mathbf{m},\quad\quad& t_\mathbf{n}^+&=\sqrt{2S}\sqrt{1-b^\dagger_\mathbf{n}b_\mathbf{n}}\cdot b_\mathbf{n}\\
        &s_\mathbf{m}^-=\sqrt{2S}\cdot a_\mathbf{m}^\dagger\sqrt{1-a_\mathbf{m}^\dagger a_\mathbf{m}},\quad\quad &t_\mathbf{n}^-&=\sqrt{2S}\cdot b_\mathbf{n}^\dagger\sqrt{1-b^\dagger_\mathbf{n}b_\mathbf{n}}\\
        &s_\mathbf{m}^z=S-a^\dagger_\mathbf{m}a_\mathbf{m},\quad\quad &t^z_\mathbf{n}&=S-b_\mathbf{n}^\dagger b_\mathbf{n}
\end{alignat}
Expanding to first order and Fourier transforming to the operators $S^\pm_\mathbf{q'}$ and $T^\pm_\mathbf{q'}$, the resulting Hamiltonian can be diagonalised by a Bogoliubov transformation to bosonic operators $\alpha_\mathbf{q'}$ and $\beta_\mathbf{q'}$ via:
\begin{equation}
    \begin{aligned}
        S^+_\mathbf{q'}=&u_\mathbf{q'}\alpha_\mathbf{q'}+v_\mathbf{q'}\beta_\mathbf{q'}^\dagger\\
        T^-_\mathbf{q'}=&u_\mathbf{q'}\beta^\dagger_\mathbf{q'}+v_\mathbf{q'}\alpha_\mathbf{q'}
    \end{aligned},
\end{equation}
where $u_\mathbf{q'}$ and $v_\mathbf{q'}$ are real coefficients fixed by requiring that all operators satisfy the relevant commutation relations, e.g. $[S^+_{{\bf q}'}, S^-_{{\bf q}''}]=2S\delta_{\mathbf{q'},\mathbf{q''}}$ and $[a_\mathbf{m},a_\mathbf{m'}^\dagger]=\delta_{\mathbf{m},\mathbf{m'}}$. This leads to the final Hamiltonian
\begin{equation}\label{eq:Hamiltonian2}
    \mathcal{H}=E_0 + \sum_\mathbf{q'}  \hbar\omega_{\mathbf{q'},0}\left(\alpha_\mathbf{q'}^\dagger\alpha_\mathbf{q'}+\frac{1}{2}\right) + \sum_\mathbf{q'} \hbar\omega_{\mathbf{q'},1}\left(\beta_\mathbf{q'}^\dagger\beta_\mathbf{q'}+\frac{1}{2}\right) ,
\end{equation}
where $E_0$ is the energy of the approximate ground state.
$\hbar\omega_{\mathbf{q}',a}$ denote the two spin wave dispersion relations that read \cite{Marshall:1971}:
\begin{align}\label{eq:AFdispersion}
    \hbar\omega^{(\rm AFM)}_{\mathbf{q'},a}=&\Omega_\mathbf{q'}\pm g\mu_BB,
\end{align} 
where $a=0,1$ is the mode index, which refers to the positive and negative signs of the Zeeman term, respectively, and
\begin{eqnarray}
    \Omega_\mathbf{q'}=\sqrt{(S(J(\mathbf{0})-(J_1(\mathbf{0})-J_1(\mathbf{q'})))-(2S-1)D)^2-J(\mathbf{q'})^2},
\end{eqnarray}
where the Fourier transformed coupling constants are again given by equations (\ref{eq:J_q}) and (\ref{eq:J1_q}).

The partial differential cross section for unpolarised neutron scattering from single magnons in an antiferromagnet is \cite{Marshall:1971}:
\begin{align}\label{eq:inel_cross_AFM}
    \left(\frac{d^2\sigma}{d\Omega dE_f}\right)_{\pm,a}=&(\gamma r_0)^2\Big[\frac{g}{2}F(\mathbf{q})\Big]^2e^{-2W}\frac{k_{\rm f}}{k_{\rm i}}(1+\hat{q}_z^2)\frac{(2\pi)^3}{4v_0}\nonumber\\&\times \sum_{\mathbf{q'},\pmb{\tau}} \left(n_{\mathbf{q'},a}+\frac{1}{2}\pm \frac{1}{2}\right)\delta(\mathbf{q}\mp\mathbf{q'}-\pmb{\tau})\delta(\hbar\omega\mp\hbar   \omega^{(\rm AFM)}_{\mathbf{q'},a})\nonumber\\&\times\big[u_\mathbf{q'}^2+v_\mathbf{q'}^2+2 u_\mathbf{q'}v_\mathbf{q'} \cos(\pmb{\rho}\cdot\pmb{\tau})\big],
\end{align}
The factor in the last line of the equation is denoted the {\em coherence factor} and consists of the coefficients of the Bogoliubov transformation. The vector $\pmb{\rho}$ connects nearest-neighbour spins on different sublattices, and is given by $\pmb{\rho}=\frac{1}{2}(a,b,c)$ for the body centered structure. 
One can show that given the condition for conservation of crystal momentum $\mathbf{q}=\pm\mathbf{q'}+\pmb{\tau}$ the following identity holds
\begin{eqnarray}
    2u_\mathbf{q'}v_\mathbf{q'}\cos(\pmb{\rho}\cdot\pmb{\tau})=2u_\mathbf{q}v_\mathbf{q},
\end{eqnarray}
meaning that the coherence factor can be calculated directly from the neutron scattering vector $\mathbf{q}$. Using this, the coherence factor is given by:
\begin{equation}
    u_\mathbf{q}^2+v_\mathbf{q}^2+2u_\mathbf{q}v_\mathbf{q}=2S \, \frac{S(J(\mathbf{0})-J(\mathbf{q})-J_1(\mathbf{0})+J_1(\mathbf{q}))-(2S-1)D}{\Omega_\mathbf{q}}.
\end{equation}
Since $J(\mathbf{q'})$ has a period of two Brillouin zones, as can be seen from equation (\ref{eq:J_q}), the coherence factor causes the scattered intensity to vary between different Brillouin zones, having the highest value when {\bf q'} is close to an antiferromagnetic ordering vector. 

\section{Implementation}
We now describe the calculations implemented in the new component \verb|SpinWave_BCO| in McStas. This component simulates the spin wave spectrum of the magnetic systems described above, incorporating both the dispersion relations eq. (\ref{eq:FMdispersion}) and eq. (\ref{eq:AFdispersion}) as well as the scattered intensity, which will be derived from equations (\ref{eq:inel_cross_FM}) and (\ref{eq:inel_cross_AFM}).

McStas simulates neutron scattering experiments using Monte Carlo ray-tracing techniques. Neutrons are treated as classical rays defined by the components of their position, velocity and spin.
To economize computing power, each ray is treated as a number of neutrons, with the equivalent neutron number given by a (non-integer) weight factor, $p$, assigned to each ray \cite{Willendrup2020}. When a ray interacts with a component, the ray parameters, including its weight, are subject to change. The weight factor is modified by the weight multiplier $w$, such that the weight factor after the interaction with component $j$ is given by 
\begin{equation}
    p_{j}=w_{j}p_{j-1}.
\end{equation}
Scattering events are simulated using Monte Carlo sampling, where we denote the sampling probability of a given event, $a$, by $f_{\rm MC}^a$. The weight multiplier of the given event, $w^a$, connects the physical probability of that event, $P^a$, to $f_{\rm MC}^a$ via the relation \cite{Willendrup2021}
\begin{eqnarray}
    P^a = w^a f_{\rm MC}^a.
\end{eqnarray}
For the \verb|SpinWave_BCO| component, the Monte Carlo choices consist of a) selecting a scattering direction, $\hat{\bf k}_{\rm f}$ and (for the antiferromagnetic system) b) selecting one spin wave mode out of $n_m=2$ possibilities. The former is determined with a uniform distribution within a predefined solid angle interval, $\Delta\Omega$. When a neutron scatters from an excitation following a dispersion relation $\hbar\omega_{\mathbf{q}}$, in this case a magnon in the chosen magnetic system, it must obey the kinematic constraint $|\hbar\omega|=\hbar\omega_{\mathbf{q}}$; fulfilling one of the delta-functions in eq. (\ref{eq:inel_cross_FM}) or eq. (\ref{eq:inel_cross_AFM}). Given that the direction of ${\bf k}_{\rm f}$ is already chosen, the kinematic constraint can be solved, giving $n_{\rm f} \geq 1$ possible values of the final neutron speed $v_{\rm f}$ \cite{squires:1977}. These values of $v_{\rm f}$ are determined, and a third MC choice c) selects one of these values. This algorithm is essentially equal to MC simulation of phonon scattering, implemented in McStas through the components \verb+Phonon_simple+ and \verb+Phonon_PG+ and described in Ref.~\cite{Davidsen2026}. It is here derived that the weight multiplier of the component is given by
\begin{equation}
    w=\frac{n_{\rm f}n_m\Delta\Omega l_{max}}{Nv_0}\frac{d\sigma}{d\Omega},
\end{equation}
where $l_{max}$ is the path length of that particular neutron ray inside the sample if it did not scatter, and $Nv_0$ is the total volume of the sample.

$d\sigma/d\Omega$ is the differential cross section, defined as
\begin{equation} \label{eq:diff_cross_sect}
\frac{d\sigma}{d\Omega} = \frac{{\rm Intensity\, of\, neutrons\, scattered\, into\, } d\Omega}{\Psi d\Omega}
\end{equation}
where $\Psi$ is the flux of the incoming neutron beam. 
The cross section thus describes the total number of neutrons scattered into a particular direction, regardless of their energy \cite{boothroyd_principles_2020}.

$d\sigma/d\Omega$ is calculated from $d^2\sigma/(d\Omega dE_{\rm f})$ by integrating over $E_{\rm f}$. When doing this, $\mathbf{v}_{\rm i}$ and $\mathbf{\hat{v}}_{\rm f}$ are kept fixed, while the neutron scattering vector and energy transfer are defined by the final neutron speed according to
\begin{align}
    \mathbf{q}(v_{\rm f})=&\frac{m_n}{\hbar}(\mathbf{v}_{\rm i}-v_{\rm f}\mathbf{\hat{v}}_{\rm f}) \\
    \hbar\omega(v_{\rm f})=& \frac{m_n}{2}(v_{\rm i}^2 -v_{\rm f}^2).
\end{align}
For details of this procedure, see Ref.~\cite{Davidsen2026}.
To integrate the inelastic cross section (\ref{eq:inel_cross_AFM}) over $E_{\rm f}$, we utilize the relation 
\begin{equation}
    \int^\infty_{-\infty}dx\:g(x)\delta(f(x))=\sum_{j=1}^{n_{\rm f}}\frac{g(x_j)}{\big|\frac{df}{dx}(x_j)\big|}=\sum_{j=1}^{n_{\rm f}}\frac{g(x_j)}{|\mathcal{J}(x_j)|},
\end{equation}
Here, $x_j$ is the $j$'th (out of $n_{\rm f}$) root of $f(x)$, and we have defined $\mathcal{J}(x) = |df/dx(x_j)|$. We can now perform the integration of the inelastic cross section, leading to the expression for the differential cross section for the creation (annihilation) of a single magnon with final speed $v_{{\rm f},j}$. For the ferromagnetic case, this leads to
\begin{align}\label{eq:diff_cross_FM}
    \left(\frac{d\sigma}{d\Omega}\right)_{\pm,j}=&  (\gamma r_0)^2\Big[\frac{g}{2}F(\mathbf{q})\Big]^2e^{-2W}\frac{{v_{{\rm f},j}}}{v_{\rm i}}(1+(\hat{q}_j^z)^2)\frac{N}{2}S\nonumber\\&\times\left(n_{\mathbf{q}_j}+\frac{1}{2}\pm\frac{1}{2}\right)\frac{m_n v_{{\rm f},j}}{|\mathcal{J}(v_{{\rm f},j})|},
\end{align}
while for the antiferromagnetic case, the expression for mode $a$ is
\begin{align}\label{eq:diff_cross}
    \left(\frac{d\sigma}{d\Omega}\right)_{\pm,a,j}=&  (\gamma r_0)^2\Big[\frac{g}{2}F(\mathbf{q})\Big]^2e^{-2W}\frac{{v_{{\rm f},j}}}{v_{\rm i}}(1+(\hat{q}_j^z)^2)\frac{N}{4}\nonumber\\&\times\left(n_{\mathbf{q}_j,a}+\frac{1}{2}\pm\frac{1}{2}\right)(u_{\mathbf{q}_j}^2+v_{\mathbf{q}_j}^2+2u_{\mathbf{q}_j}v_{\mathbf{q}_j})\frac{m_n v_{{\rm f},j}}{|\mathcal{J}_a(v_{{\rm f},j})|}.
\end{align}
In the simulation, $\mathcal{J}(v_{\rm f})=\frac{\partial}{\partial v_{\rm f}}(|\hbar\omega(v_{\rm f})|-\hbar\omega_{\mathbf{q}(v_{\rm f})})$ is found numerically for the selected value of $v_{\rm f}$.

The input parameters for the component are listed in table \ref{tab:comp_input_parms}.
\begin{table}[H]
    \centering
    \caption{Table of input parameters for the spin wave component SpinWave\_BCO.}
    \begin{tabular}{|c|c|}
        \hline
        Parameter name & Description \\
        \hline\hline
        radius, yheight [m]& Radius and height of cylindrical sample.\\\hline
        sigma\_inc, sigma\_abs [barns]&Cross sections for incoherent scattering and absorption.\\\hline
        target\_index& \makecell{Relative index of component to focus at \\with next component being +1.}\\\hline
        target\_x, target\_y, target\_z [m] &Coordinates to focus at (alternative to target\_index).\\\hline
        focus\_xw, focus\_yh [m]&Width and height of focusing.\\\hline
        focus\_aw, focus\_ah [rad]&Direct specification of $d\Omega$.\\\hline
        FM &\makecell{ Flag for type of order. Default is 0 (antiferromagnet), \\1 selects ferromagnetic.}\\\hline
        a, b, c [Å]& Lattice constants of orthorhombic lattice unit cell.\\\hline
        S & Spin of magnetic ions.\\\hline
        j [meV]& \makecell{Magnetic interaction constant to the 8 equivalent \\neighbours in the body centered positions.}\\\hline
        j\_a, j\_b, j\_c [meV]&\makecell{Magnetic interaction constants to the neighbours\\ along the $a$-, $b$-, and $c$-axes, respectively.}\\\hline
        j\_110, j\_110\_prime [meV]&\makecell{Magnetic interaction constants for altermagnetic splitting\\ when FM=0 (see section \ref{sec:alter}). Both are 0 by default.}\\\hline
        D [meV]& Uniaxial single-ion anisotropy constant.\\\hline
        B [T]&Magnetic field strength.\\\hline
        T [K]&Temperature.\\\hline
        mode\_input& \makecell{If FM=0: Parameter to specify which mode is simulated.\\ Value of 2 (default) simulates both modes. \\Values of 0 or 1 simulates only the corresponding mode.}\\
        \hline
    \end{tabular}
    \label{tab:comp_input_parms}
\end{table}

When the component is placed without rotation, the $a$-, $b$-, and $c$-axes lie along the $x$-, $y$-, and $z$-axes of the McStas coordinate system, respectively.

The main functionality of the component is listed in Algorithm~\ref{alg:spinwave}. In these calculations, by far the largest computational power is spent finding the relevant values of $v_{\rm f}$ that simultaneously fulfil the dispersion relation and the kinematic constraint.

The component also conducts a number of checks of the calculated expressions to ensure that the input parameters lead to physically meaningful results.
An error message is produced if the dispersion becomes negative or imaginary and if the coherence factor becomes negative when simulating an antiferromagnetic system.

These would likely be caused by the input parameters that do not correspond to the chosen ground state (FM or AFM). When simulating a ferromagnetic system, the value of $B$ is forced to be positive (or zero), such that the applied field always stabilises the ground state. For the antiferromagnetic system, the magnetic field strength is set to zero if it induces a spin-flop transition, i.e. if the dispersion for the low-energy mode becomes negative.

\begin{algorithm}
\caption{Magnon Scattering Procedure} \label{alg:spinwave}
\begin{algorithmic}[1]
  \State Choose position inside sample to scatter from
  \State Choose direction within $\Delta\Omega$ to scatter into
  \If{FM = 0}
      \If{mode = 2}
        \State Choose spin wave mode
     \Else
        \State Pick given mode
     \EndIf
  \Else 
    \State Chose ferromagnetic system
  \EndIf
  \State Find possible values of $v_{\rm f}$ by solving $|\hbar\omega|=\hbar\omega_{\mathbf{q}}$ numerically
    \State Choose one of the $n_{\rm f}$ values of $v_{\rm f}$
    \State Update the neutron velocity
    \State Calculate $w$
    \State Multiply $p$ with $w$ to perform the weight transformation
\end{algorithmic}
\end{algorithm}

\section{Validation}
The component was validated by comparing theoretical calculations with simulated data.
This includes both a validation of the simulated dispersion relation as well as the simulated absolute intensities.

 The main focus was on the validation of the antiferromagnet. This was done by modeling the spin wave spectrum of MnF$_2$. This material has previously been understood as a classical antiferromagnet, with the magnetic Mn$^{2+}$ situated on a body-centered tetragonal lattice. The crystal- and Hamiltonian parameters for MnF$_2$ are listed in table \ref{tab:param_input}.

However, we will start by showing that the ferromagnet produces the result expected from theory. To do this, we artificially set all j values to be negative, keeping their magnitude.

We employ a virtual thermal Triple-Axis Spectrometer (TAS), operating with a fixed final energy of $E_{\rm f} = 14.7$~meV. 
The instrument was previously used to test the \verb|Phonon_PG| component and is described in \cite{Davidsen2026}.
The simulated instrument is tuned to have an unrealistically high resolution, by reducing the sizes of the instrument components, such as the source, monochromator, sample, and analyser. Furthermore, the simulations do not include any sources of background or elastic scattering from the crystal. All of this aids in the comparison between the simulated datasets and the linear spin wave theory results.

\begin{table}[H]
\caption{Input parameters to the Spin wave component used in the validation process. The antiferromagnetic parameters match those modelled for the compound MnF$_2$; adopted from Ref.~\protect\cite{Yamani:2010}.}
\parbox{.45\linewidth}{
\centering
    \begin{tabular}{|c|c|}
    \hline
        \multicolumn{2}{|c|}{Ferromagnet}\\
       \hline
        Parameter & Value\\
        \hline\hline
        $a=b$ & $4.873$ Å\\
        \hline
        $c$ & $3.130$ Å\\
        \hline
        $S$ & $2.5$ \\
        \hline
        $j$ & $-0.304$ meV \\
        \hline
        $j_c$ & $-0.05  6$ meV\\
        \hline
        $j_a=j_b$ & $-0.008$ meV\\
        \hline
        $D$ & $-0.023$ meV\\
        \hline
    \end{tabular}
}
\hfill
\parbox{.45\linewidth}{
\centering
    \begin{tabular}{|c|c|}
    \hline
        \multicolumn{2}{|c|}{Antiferromagnet}\\
       \hline
        Parameter & Value\\
        \hline\hline
        $a=b$ & $4.873$ Å\\
        \hline
        $c$ & $3.130$ Å\\
        \hline
        $S$ & $2.5$ \\
        \hline
        $j$ & $0.304$ meV \\
        \hline
        $j_c$ & $-0.05  6$ meV\\
        \hline
        $j_a=j_b$ & $0.008$ meV\\
        \hline
        $D$ & $-0.023$ meV\\
        \hline
    \end{tabular}
}
\label{tab:param_input}
\end{table}

The data for the simulated dispersion were obtained as constant-energy scans along four paths in {\bf q}-space, each of which was confined to the $(h 0 l)$-plane. Each scan was performed with a step size of $0.01$~meV from 0 to 8~meV for the antiferromagnet and 0 to 13.2~meV for the ferromagnet. The sample temperature was set to $T=2$~K for all simulations, much lower than the ordering temperature of MnF$_2$, $T_{\rm N} = 67$~K.

\subsection{The ferromagnetic case}
\begin{figure}[H]
    \centering
    \includegraphics[width=.95\linewidth]{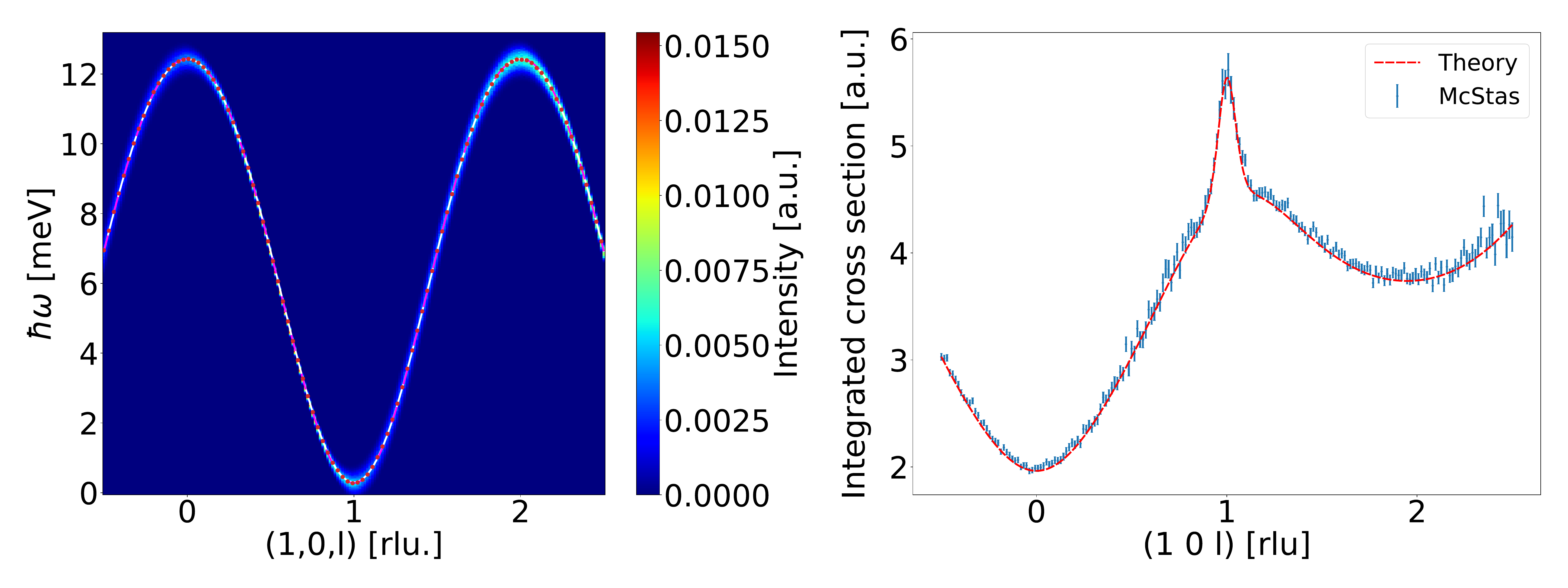}
    \caption{Simulated data for the ferromagnetic system. (left) Simulated dispersion compared with theory. Red points show the fitted peak positions. The solid white line shows the dispersion fitted to the red points and the dashed magenta line shows the dispersion calculated from theory. (right) Integrated differential cross section compared with theory.}
    \label{fig:ferromagnet_data}
\end{figure}

The ferromagnetic dispersion has been simulated along $(10l)$. A magnetic field of $B=2~T$ is applied to increase the size of the gap at the Brillouin zone center. This is done to separate the spin wave creation and annihilation signals.
A very small overlap between these two signals is still present, but the effect is relatively small. Gaussian line shapes have been fitted to the constant-$\mathbf{q}$ scans to give a rough estimate of the intensity peak centers. The analytic expression for the dispersion, eq.~(\ref{eq:FMdispersion}), is fitted to the $\mathbf{q}$-dependence of the fitted peak positions. Since the scan is along $l$, only $j_c$, $j$ and $D$ are kept as free parameters in the fit. The intensity data, fitted dispersion, and the dispersion from theory can all be seen in figure \ref{fig:ferromagnet_data} (left). The parameters of the fit are reported in table \ref{tab:fit_parms_ferro}.

\begin{table}[H]
    \centering
    \caption{Dispersion parameters obtained from fit to simulated ferromagnetic data compared to the actual values used in the simulation.}
    \begin{tabular}{|c|c||c|}
    \hline
        Parameter & Value from fit [meV]& Input value to simulation [meV]\\
        \hline
        $j$ & $-0.303184(8)$ & $-0.304$\\
        $j_c$&$-0.05659(3)$ & $-0.056$\\
        $D$&$-0.0332(1)$ & $-0.023$\\
        \hline
    \end{tabular}

    \label{tab:fit_parms_ferro}
\end{table}
Although the shape of the fitted dispersion is seen to match theory, the extracted parameters do not exactly match those used in the simulation; the value of the anisotropy constant, $D$ deviates around 50\%. However, given that $D$ is very small and mainly determined by the data near the gap, where the creation and annihilation signals overlap, it is not surprising that this analysis shows minor deviations. For the antiferromagnetic case we show how a more in-depth analysis is able to reproduce the simulation parameters with a much higher accuracy.

In addition to fitting the dispersion, we compare the simulated relative intensities with those expected from theory. 
The intensities are first normalized to the incoming flux and then integrated along the energy transfer, which also removes the instrument-dependent energy broadening. The values from theory are normalized to the $(1,0,-0.5)$ point of the data, and the comparison is shown in figure \ref{fig:ferromagnet_data} (right). The simulated data are seen to reproduce the theoretical values with excellent agreement.

\subsection{The antiferromagnetic case}

For the antiferromagnetic system, we 
simulate constant-$q$ scans from the classical antiferromagnet MnF$_2$. 

The Hamiltonian input parameters were taken from reference \cite{Yamani:2010} and are shown in table \ref{tab:param_input}. 

\begin{figure}
    \centering
    \includegraphics[width=0.3\linewidth]{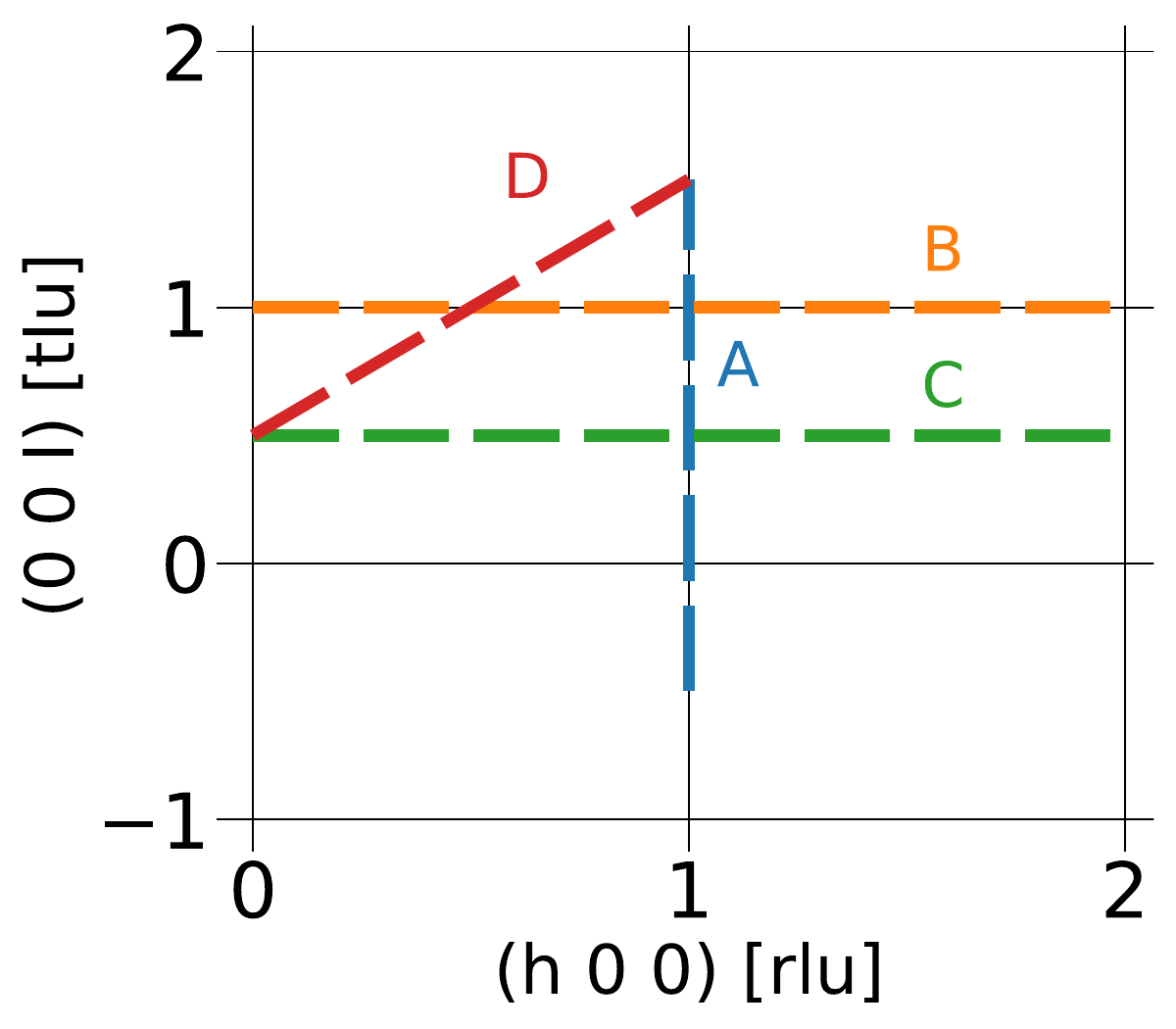}
    \caption{Map over scanned directions in reciprocal space. Labels correspond to those seen in figure \ref{fig:intensity_plots}.}
    \label{fig:rlu_map}
\end{figure}

The dispersion was simulated along four directions in reciprocal space, which is illustrated in figure \ref{fig:rlu_map}.

\begin{figure}[H]
    \centering
    \includegraphics[width=\linewidth]{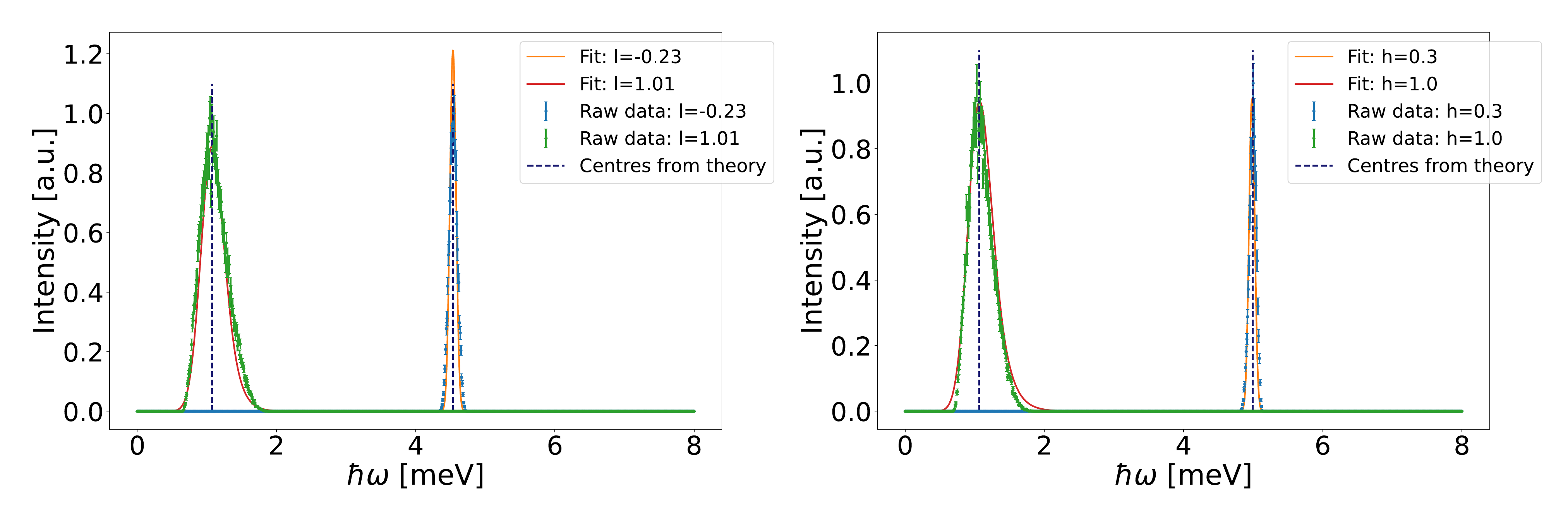}
    \caption{Examples of constant-{\bf q} scans for the antiferromagnetic case. Left is along $(10l)$ and right is along $(h01)$. The solid lines show the fitted functions used to find the intensity peak centers and the dashed lines show the peak centers from theory.}
    \label{fig:slices}
\end{figure}
As in the ferromagnetic case, the center of each constant-$\mathbf{q}$ scan intensity peak was found by fitting Gaussian line shapes to the data, as illustrated in figure \ref{fig:slices}. Due to the instrumental resolution function, the shape of the intensity peaks varies between different values of {\bf q}. A symmetric Gaussian function is used to fit most peaks. However, near the Brillouin zone centers, which is at the bottom of the dispersion, the resolution function makes the peaks highly asymmetric. For these scans, an approximate resolution convoluted-gap function for TAS instruments (described in Ref. \cite{Lenander2025}) are used to fit the mode positions. In this function, it is assumed that the {\bf q}-resolution is coarse in two out of three directions and at the gap, the dispersion follows a parabola, which are both valid in this setup. The function describes the asymmetric shape well, examples of the fits can be observed in figure \ref{fig:slices} at low energies. Due to the extremely high energy resolution of the instrument and the assumption that the energy resolution is not correlated to the {\bf q}-resolution, the fitted mode positions are slightly underestimated (deviations less than 0.1~meV).

The analytical expression for the dispersion, eq.~(\ref{eq:AFdispersion}), is fitted to the {\bf q}-dependence of the fitted peak positions. We fit all four datasets at once, where only the interaction parameters, $j, j_a, j_c$, and $D$, are allowed to vary. This gives us a direct comparison to the theoretical dispersion relation.

\begin{figure}[ht] 
    \begin{center}
    \includegraphics[width=\textwidth]{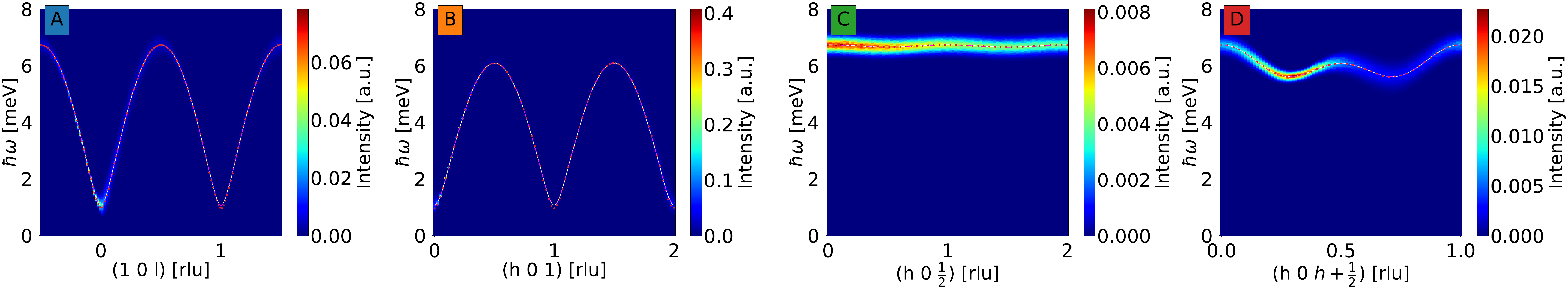}  
    \end{center}
    \caption{Simulated intensity data measured along four directions in $\mathbf{q}$-space. Points show fits to intensity peaks. The solid white line is the dispersion fitted to these points and the dashed magenta line shows the dispersion calculated from eq.~(\ref{eq:AFdispersion}).}
    \label{fig:intensity_plots}
\end{figure}

In figure \ref{fig:intensity_plots} we perform this comparison by displaying the raw intensity data, the dispersion values obtained from the constant-{\bf q} fits, the fitted dispersion relation, and the theoretical dispersion. 
 
The values of the fitted parameters are reported in table~\ref{tab:fit_parms}, with the statistical uncertainties from the fits. Compared with the actual interaction values used as input to the component, we see that the general agreement is excellent, with the largest deviation being only around 2.5\% (on the small $D$ parameter). This shows that our simulation method produces the correct dispersion and is self-consistent.

\begin{table}[H]
    \centering
    \caption{Dispersion parameters obtained from fitting the simulated antiferromagnetic data compared to the actual values used in the simulation.}
    \begin{tabular}{|c|c||c|}
    \hline
        Parameter & Value from fit [meV]& Input value to simulation [meV]\\
        \hline
        $j$ & $0.30372(1)$ & $0.304$\\
        $j_a$&$0.00800(2)$ & $0.008$\\
        $j_c$&$-0.05713(2)$ & $-0.056$\\
        $D$&$-0.02359(1)$ & $-0.023$\\
        \hline
    \end{tabular}

    \label{tab:fit_parms}
\end{table}
We will further validate the intensities simulated by the component used for the case of MnF$_2$. This is done by recording the simulated differential cross section, obtained through eq. (\ref{eq:diff_cross_sect}), at several points on the dispersion curve. In the simulations, we have directly recorded the incoming flux and the outgoing intensity by placing energy-sensitive monitors just before the sample and analyser in the TAS setup. The solid angle was calculated from known geometry parameters.

We compare these values to the results of spin wave theory through eq.~(\ref{eq:diff_cross}), as shown in table \ref{tab:abs_inten}. 
We see that the agreement is generally excellent. The largest relative deviation is around 10\% for the very weak annihilation signal; $h\omega = -1.06$~meV. The deviation is around 5\% near the low-intensity ferromagnetic point (1 0 1). All other deviations are on the order of 2\% or lower.
\begin{table}[ht] \small
    \centering
    \caption{Simulated (''sim``) differential cross sections compared with results calculated from the analytic (''ana``) expression at different points on the dispersion, defined by the point $(\mathbf{q},\hbar\omega)$ and index of the simulated mode.}
    \begin{tabular}{|c|c|c|c|c|c||c|c|c||c|}
    \hline
    $h$&$k$&$l$&$\hbar\omega$ [meV]&Mode&B [T]&$\Psi_\text{mon}$ [cm$^{-2}$s$^{-1}$]&$I_\text{det}$ [s$^{-1}$]&$\frac{d\sigma}{d\Omega}$(sim) [cm$^2$]&$\frac{d\sigma}{d\Omega }$(ana) [cm$^2$]\\
    \hline 
        1 & 0 & 0 &$1.06$&0&0&$3.509\cdot10^{5}$&$3.750\cdot10^{-1}$&$1.069\cdot10^{-2}$& $1.074\cdot10^{-2}$\\
        1&0&0&-1.06&0&0&$2.790\cdot10^4$&$5.997\cdot10^{-5}$&$2.149\cdot10^{-5}$&$2.453\cdot10^{-5}$  \\
        1 & 0& 1& $1.06$&0&0&$3.535\cdot10^{5}$&$5.162\cdot10^{-3}$&$1.460\cdot10^{-4}$& $1.378\cdot10^{-4}$ \\
        1&0&0.5 &6.73&0&0&$5.518\cdot10^5$&$6.193\cdot10^{-2}$&$1.122\cdot10^{-3}$& $1.097\cdot10^{-3}$ \\
        0.5&0&0.5&6.65&0&0&$5.472\cdot10^5$&$7.499\cdot10^{-2}$& $1.370\cdot10^{-3}$& $1.363\cdot10^{-3}$\\
        2&0&0.3&5.48&0&0&$5.070\cdot10^5$&$1.550\cdot10^{-2}$&$3.069\cdot10^{-4}$& $3.030\cdot10^{-4}$ \\
        0.7&0&0.1&5.15&0&0&$4.957\cdot10^5$&$5.472\cdot10^{-2}$&$1.104\cdot10^{-3}$& $1.085\cdot10^{-3}$\\ 
        
        1&0&0&1.29&0&2&$3.609\cdot10^5$&$3.882\cdot10^{-1}$&$1.076\cdot10^{-2}$& $1.065\cdot10^{-2}$\\
        1&0&0&0.83&1&2&$3.472\cdot10^5$&$3.795\cdot10^{-1}$&$1.093\cdot10^{-2}$&$1.089\cdot10^{-2}$\\
        \hline
    \end{tabular}

    \label{tab:abs_inten}
\end{table}

The simulated intensities are also compared to a simulation of the dynamical correlation function performed using the MATLAB library SpinW \cite{Toth:2015}. The datasets measured along the $(1 0 l)$ direction are used for this. To compare the two simulated datasets, each is converted to quantities proportional to the inelastic cross section. The McStas intensities are normalized and integrated along the energy transfer, as was done for the ferromagnetic data.

\begin{figure}
    \centering
    \includegraphics[width=0.75\linewidth]{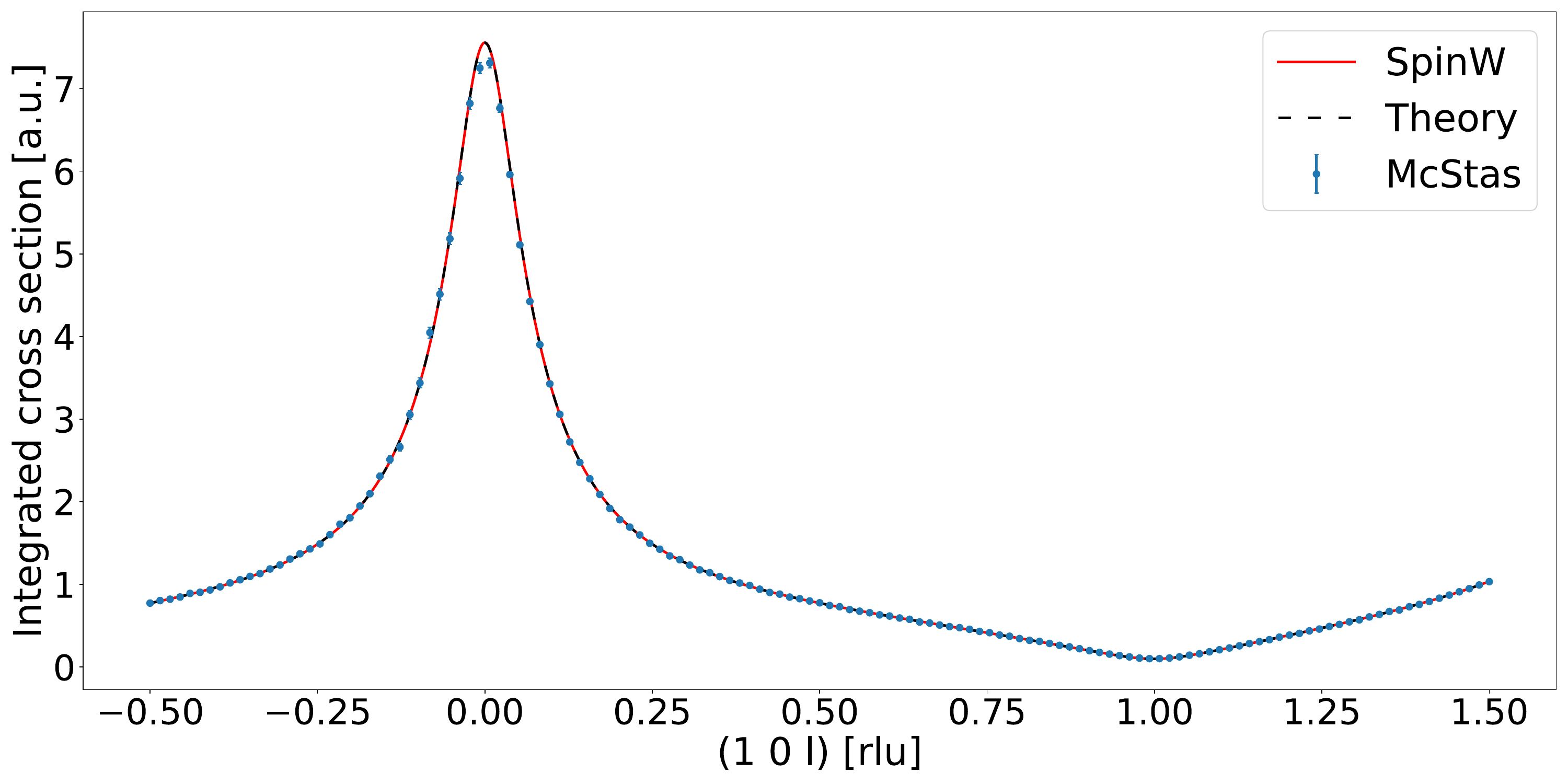}
    \caption{Comparison of integrated inelastic cross section between McStas and SpinW. Data has been normalised to the (1 0 $\frac{1}{2}$) point.}
    \label{fig:McStas_v_SpinW}
\end{figure}

The SpinW data is multiplied by the instrument dependent factor $k_f/k_i$, and this is then normalized to the $(1,0 ,-0.5)$ point of the McStas data for comparison. The two datasets can be seen in figure \ref{fig:McStas_v_SpinW}. An excellent agreement is found, showing that the \verb|SpinWave_BCO| component is able, within a constant scaling factor, to accurately reproduce results obtained by SpinW.

The difference between the direct measurement of the differential cross section in table \ref{tab:abs_inten} and the integrated cross section in figure \ref{fig:McStas_v_SpinW} are a clear demonstration of the two distinctly different ways to integrate through reciprocal space: The differential cross section is an integration over $E_{\rm f}$ while keeping the scattering angle fixed, leading to the Jacobian, while figure \ref{fig:McStas_v_SpinW} shows an integration over $E_{\rm f}$, keeping $\mathbf{q}$ fixed, leading to data which is symmetric within a given Brillouin zone. As we demonstrate these two equivalent methods both yield accurate results.

\begin{figure}
    \centering
    \includegraphics[width=0.9\linewidth]{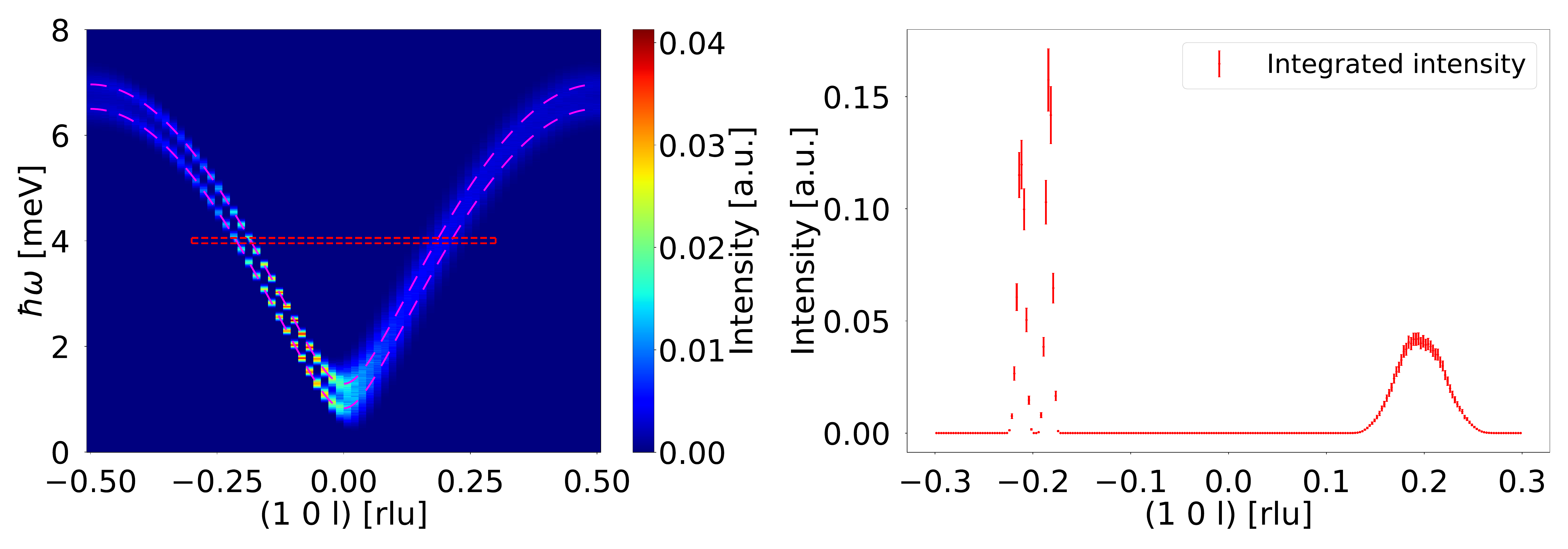}
    \caption{Result of simulations made with an applied magnetic field of $B=2$ T. (left) The full $({\bf q}, \hbar\omega)$ map showing all simulation data. (right) A cut through this data at $\hbar\omega = 4$~meV and integrated along energy transfer with $\Delta E=\pm0.055$ meV (red dashed line in the left panel).}
    \label{fig:magnetic}
\end{figure}

To further demonstrate the capabilities of the component, a simulation is performed with an applied magnetic field of $B=2$ T. The data is shown in figure \ref{fig:magnetic}. The two modes are split by the magnetic field, as expected. The magenta lines show the calculated spin wave mode dispersions, showing that the simulated data again matches well with theory.
The area within the red outline was simulated using a {\bf q}-spacing six times smaller than the rest of the data. This data is shown integrated along the energy transfer on the right in figure \ref{fig:magnetic}. This clearly shows how the splitting is not always resolved. This is due to the orientation of the resolution ellipsoid, which is significant even for our high-resolution TAS \cite{Shirane:2002}.

\section{Altermagnetism in \texorpdfstring{MnF$_2$}{MnF2}} \label{sec:alter}
As an additional functionality, we have implemented a simple altermagnetic dispersion. Altermagnetism in MnF$_2$ was suggested in Ref.~\cite{Smejkal2022}. While previous searches for the altermagnetic splitting of the magnon modes in MnF$_2$ were unsuccessfully \cite{Morano2025}, a recent study revealed this splitting using polarised neutron scattering \cite{McClarty:altermagnetisme_revealed}. In the latter study, the spin-wave spectrum was modelled using both altermagnetic exchange couplings and long-ranged dipolar couplings. We will use a simpler model, only including the altermagnetic couplings, to show how the \verb|SpinWave_BCO| component can be used for this purpose.

\begin{figure}[H]
    \centering
    \includegraphics[width=0.6\linewidth]{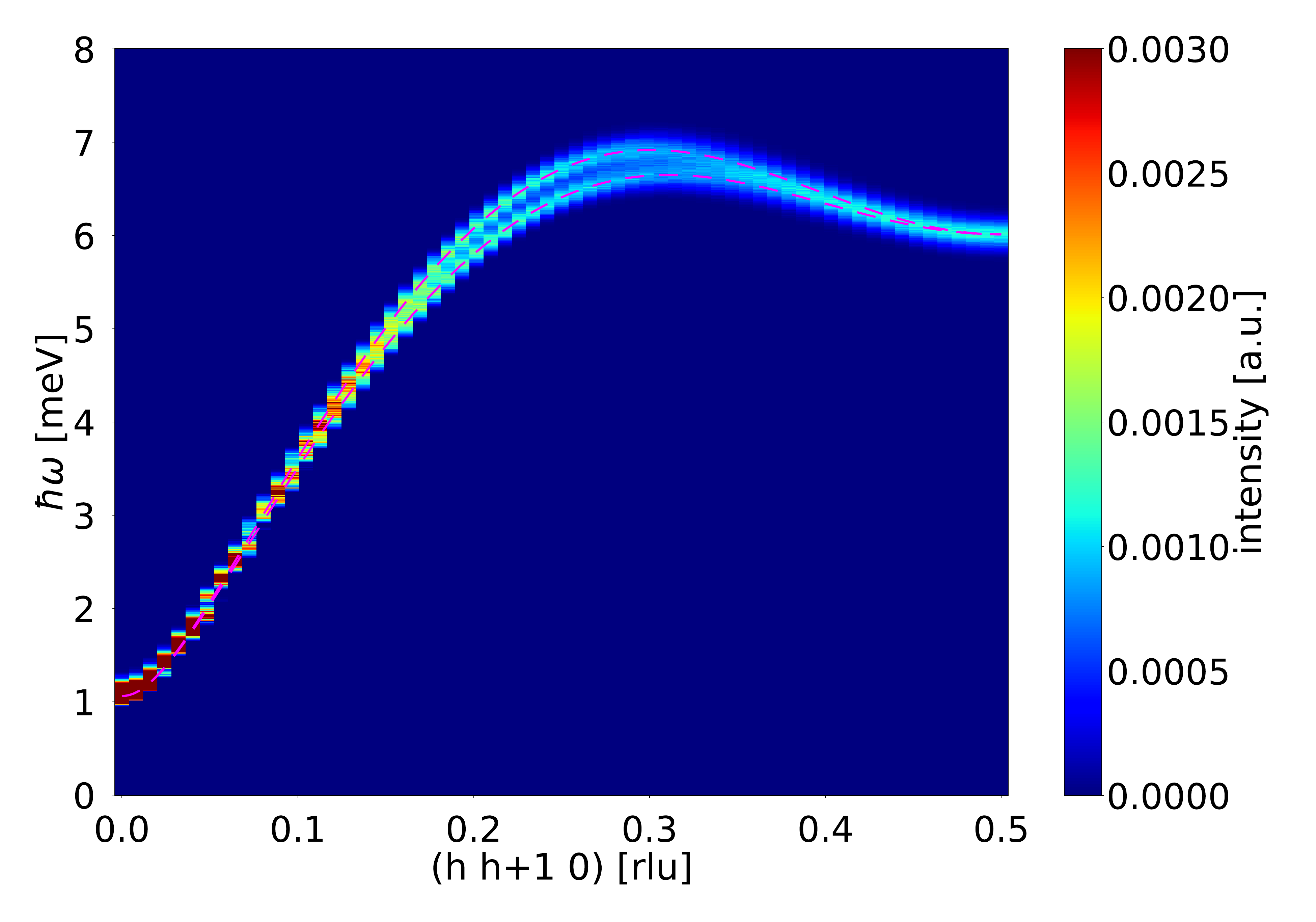}
    \caption{Simulated altermagnetic dispersion along $(h,h+1,0)$. The dashed magenta lines show the theoretical dispersion relations for the split spin wave modes.}
    \label{fig:altermagnet}
\end{figure}

The theory of exchange-driven altermagnetic spin waves in MnF$_2$ is described in \cite{McClarty_altermagnetism}. The lowest order exchange interactions which lead to altermagnetic ordering are those between Mn$^{2+}$ ions on the same sublattice at positions characterized by the vectors $\mathbf{\hat{x}}+\mathbf{\hat{y}}$ and $\mathbf{\hat{x}}-\mathbf{\hat{y}}$, respectively. These coupling constants will be denoted $j_{110}$ and $j_{110}'$, respectively. Including these in the Hamiltonian and performing the same diagonalization procedure as for the antiferromagnetic case leads to the altermagnetic dispersion relation. First, the Fourier transform of the coupling constants within each sublattice reads
\begin{equation}
    J_1^{(\rm AM)}(\mathbf{q'})=J_1(\mathbf{q'})+2(J_{110}+J'_{110})\cos\left(aq_x'\right)\cos\left(bq_y'\right)
\end{equation}
with $J_(\mathbf{q'})$ given by equation (\ref{eq:J1_q}). Using the new $J_1^{(\rm AM)}(\mathbf{q'})$, the spin wave dispersion becomes:
\begin{equation} \label{eq:alt_disp}
    \hbar\omega^{(\rm AM)}_{q',a}=\Omega_\mathbf{q'}\pm(2(J_{110}-J'_{110})\sin\left(aq_x'\right)\sin\left(bq_y'\right)+g\mu_BB)
\end{equation}
For $J_{110}=J'_{110}$, this reduces to the known antiferromagnetic result. For $J_{110}\neq J'_{110}$ the two modes are split by the extra $\mathbf{q'}$-dependent term. This term is essentially added to the magnetic field, and just as the Bogoliubov transformation and the coherence factor are independent of the field, they are also independent of this term. Therefore, the intensity is essentially unchanged by the altermagnetic splitting. For this reason, we have implemented the altermagnetic splitting by simply modifying the dispersion relation according to eq.~(\ref{eq:alt_disp}). A simulation of the altermagnetic dispersion has been performed on the same TAS setup as used before. To show the splitting, very large values of $j_{110}$ and $j_{110}'$ have been used: $j_{110}=-0.25\cdot j$ and $j_{110}'=-0.5\cdot j$. The simulated spectrum can be seen in figure \ref{fig:altermagnet}, where the theoretical dispersion relation has been plotted for comparison. The altermagnetic splitting is clearly seen in the data.

\section{Discussion and conclusion}
We have demonstrated the implementation of accurate magnetic dynamics in McStas with a new sample component \verb|SpinWave_BCO|, which simulates inelastic neutron scattering from spin waves in a ferromagnet or a two-sublattice antiferromagnet in a body-centered orthorhombic crystal structure. In addition, we have implemented and tested an altermagnetic splitting of the antiferromagnetic spin wave dispersion.
 
The simulated dispersion relation and differential cross section are found from linear spin wave theory. Simulated data has been compared directly to theory and to the program SpinW to verify the McStas simulations. This was exemplified by simulating the antiferromagnetic spin wave spectrum of MnF$_2$.

We find that the simulated dispersions and absolute intensities agree extremely well with linear spin wave theory, both for magnon creation and annihilation. This has been tested for many different values of {\bf q}. 

While our present work is an important step towards the description of magnetic dynamics in McStas, much development work lies ahead in this direction. The feature of greatest relevance will surely be to expand the description of lattice geometries and interactions to be able to model all crystal classes and also accommodate spin waves from $n$-sublattice systems and spiral order, in analogy with SpinW \cite{Toth:2015}. To promote the calculation of realistic intensities, it could be of relevance to include a library of the magnetic structure factors, $F(q)$, for the most common magnetic ions. A very ambitious, but also relevant, update would be to include a full description of neutron polarization, which is generally accommodated in McStas \cite{Knudsen:2014}, but the effect of which is not included in the \verb|SpinWave_BCO| component.

\begin{acknowledgements}
We thank Rasmus Toft-Petersen for expressing the urgency with which this spin wave component was developed and Henrik M.~R\o nnow for suggesting to include the altermagnetic dispersion. Furthermore, we thank Amalie F.\ Davidsen and Frida B.\ Nielsen for useful discussions on inelastic scattering in McStas. We also thank Jakob Lass for supplying the SpinW simulations of MnF$_2$ used to validate the McStas simulations. Finally, we thank Peter Willendrup and Mads Bertelsen for including the \verb|SpinWave_BCO| component into McStas' component library.
\end{acknowledgements}

\begin{funding}
This project was supported by the Danish Committee for Research Infrastructure (NUFI) through funding to the ''ESS-Lighthouse`` Q-MAT.
\end{funding}

\ConflictsOfInterest{The authors declare no conflicts of interest.
}

\DataAvailability{All simulation files, simulated data, and scripts for data analysis can be obtained from the authors upon reasonable request.
}

\bibliography{iucr} 

\end{document}